# A large amplitude flaring dMe star in the 1978 October 6B $\gamma$-ray burst error box *

**J. Greiner[1] and C. Motch[2]**

[1] Max-Planck-Institut für Extraterrestrische Physik, D-85740 Garching, Germany
[2] CNRS, UA 1280, Observatoire de Strasbourg, 11 rue de l'Université, F-67000 Strasbourg, France

**Abstract.** Spectroscopic observations provide evidence that the optical transient object observed in 1966 on archival plates of Sonneberg Observatory is a B=18.5 dMe star. For an assumed flare duration of 40 min. (or longer) the flare amplitude was $\triangle m_B \approx 5.2$ mag. If the recorded event was belonging to the class of 'fast' flares its amplitude could have reached 7-9 mag.. We estimate a lower limit for the optical flare energy release of $10^{35}$ $(d/100$ pc$)^2$ erg/s.

We discuss the positional coincidence of this large amplitude flaring star with the error box of the $\gamma$-ray burst 1978 October 6B and conclude that the flare star is probably not related to the $\gamma$-ray burst. The fact that no other optical flare was found from this object restricts their frequency to be less than $7 \times 10^{-4}$ h$^{-1}$ for amplitudes (in B) greater than 4.5 mag. Nevertheless, large amplitude flares constitute a serious background problem for the wide field search for possible optical counterparts of $\gamma$-ray bursters.

## 1. Introduction

During a search for flaring optical counterparts to $\gamma$-ray burst sources (GRBs) on archival patrol plates, an optical outburst image was found in the 3200 arcmin$^2$ error box of the GRB which occured on October 6, 1978 (GB 781006B = GBS 0008+13; Greiner, Naumann & Wenzel 1991) and was named S 10933 according to the convention of new Sonneberg variables. This outburst image appears on three simultaneously exposed plates taken on 1966 August 14/15 from 23.33 to 0.13 UT. These consist of a pair of a blue and red plate of the same field and another blue plate of the neighbouring field where the object is just at the edge of the plate (on the corresponding red plate of this neighbouring field the object is invisible due to vignetting and general lower limiting magnitude of the red plates). The brightness estimates yielded $m_{pg} = 13\overset{m}{.}3 \pm 0.1$ mag and $m_v = 13\overset{m}{.}8 \pm 0.2$ mag with a high negative colour index of C = $- 0\overset{m}{.}5 \pm 0.3$ mag.

The optical transient object is invisible on a plate taken during the preceding night (rise of $> 1.7$ mag in $< 1$ day) and on September 10 (decline of $> 1.7$ mag in $< 27$ days) (both plates are of good quality with a threshold magnitude of about $15^{th}$ mag pg); the overall brightness range would be $> 4$–5 mag pg.

The coordinates of these images (measured on the Tessar plate Te$_3$ 5154) are (equinox 1950.0) $\alpha = 0^h 10^m 07\overset{s}{.}6$, $\delta = + 12° 51' 24''$ ($1\sigma$ error of $\pm 2''$). Within the $3\sigma$ error circle of the position of the object a star of $18^{th}$ mag in the blue band is visible on the Palomar Observatory Sky Survey prints. Here, we report spectroscopic observations of this faint object.

## 2. Observations

### 2.1. Imagery

CCD B and R images of the field containing the optical transient were obtained from 1991 November 10 to 12 with the 1.2m telescope at Observatoire de Haute-Provence (OHP). The Cassegrain focus camera was equipped with a RCA SID 501EX CCD chip yielding a pixel size of $0\overset{''}{.}86$ on the sky. A total of 20 min in B and 70 min in R band were obtained during the 3 nights of observation. Raw CCD frames were corrected for flat-field and bias using standard MIDAS procedures. We show in Fig. 1 and 2 the mean corrected R and B images.

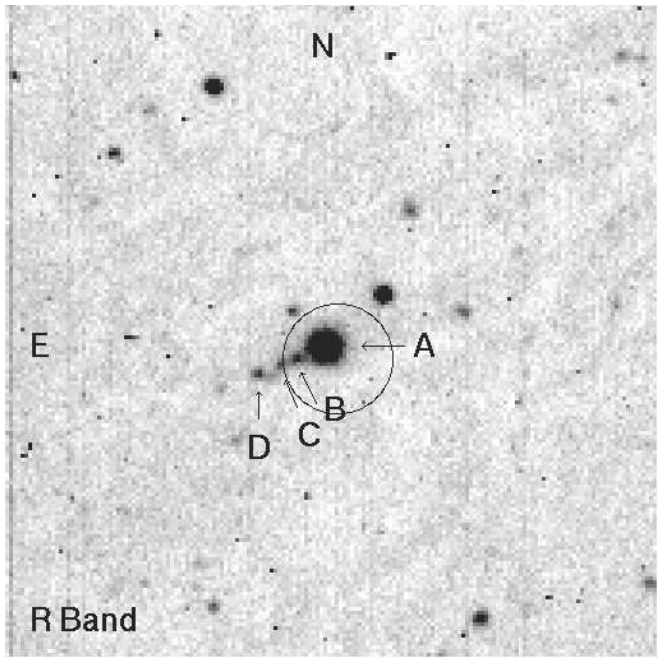

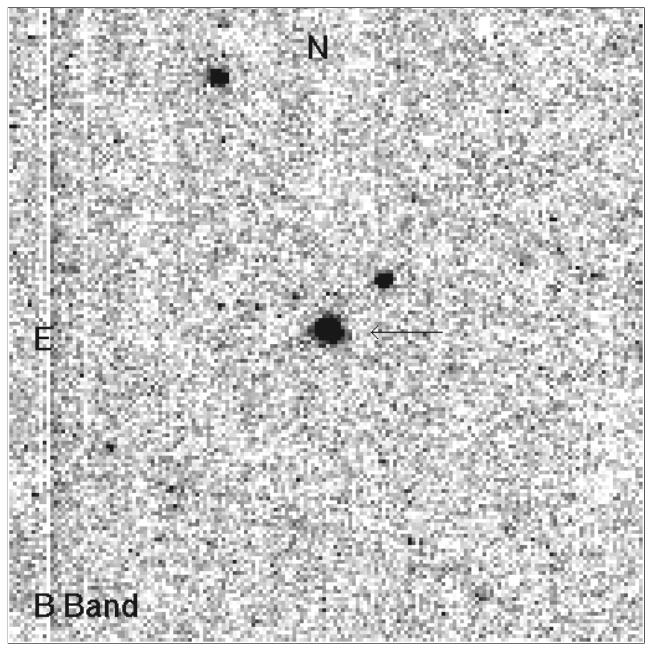

**Fig. 1.** Sum of all R band images corresponding to a total exposure time of 70 min. The position of the $3\sigma$ error circle of the optical transient recorded in 1966 is shown for comparison. The error radius ($12\farcs6$) includes the uncertainty on the astrometric calibration of the CCD frame. Object A is the dMe star probably responsible for the optical flash. The structure (B,C,D) extending eastwards from the flare star consists of several faint stars accidentally aligned on the line of sight.

**Fig. 2.** B band image corresponding to a total exposure time of 20 min.

*2.2. Spectroscopy*

Spectroscopic observations were carried out on two occasions. Low and medium resolution spectra of the brightest optical object A (see Fig. 1) were obtained at OHP on 1992 February 4,7 and 8. For these observations we used the CARELEC spectrograph attached at the 1.93m telescope and a CCD TK 512 chip. At low resolution, the spectral range was 3800–7300 Å with a FWHM resolution of 17 Å. One medium resolution spectrum ($\lambda\lambda$ 4100–4900 Å; FWHM resolution 3.5 Å) was acquired on February 7.

Further spectroscopic observation were obtained using the 3.5 m telescope at Calar Alto on Sep 28 – Oct 1, 1992. We used the Cassegrain spectrograph with a 1024×640 RCA chip (pixel size 15 $\mu$m). At low resolution, a grating with 240 Å/mm was used, allowing the range 3800–7200 Å to be observed with a FWHM resolution of 10 Å. The slit was put parallel to the row of faint objects apparently aligned with the bright candidate (see Fig. 1). On Sep 30, 1992 three spectra with a grating of 60 Å/mm (corresponding to 0.9 Å/pixel) were acquired in the range $\lambda\lambda$ 6200–7100 Å (FWHM resolution 2.5 Å).

All long slit spectra were corrected for bias and flatfield and calibrated in wavelength using standard MIDAS reduction packages.

### 3. The content of the optical transient error circle

*3.1. Central object A*

Fig. 3 shows the mean low resolution spectrum of the bright candidate star obtained at Calar Alto. The spectral type was estimated by comparing our flux calibrated spectra with spectra in electronic atlases (Jacoby, Hunter and Christian (1984) and Turnshek et al. (1985)). TiO bands indicate a spectral type later than M2 while the CaOH band at $\lambda\lambda$ 5530–5570 suggests a spectral type earlier than M6. Balmer lines are clearly seen in emission up to H$\gamma$ and the CaII H&K emission lines are also detected. H$\alpha$ equivalent width varies from 7 to 13 Å (see Table 1). Finally, the strength of the NaI doublet ($\lambda\lambda$ 5890–96) indicates a dwarf star. We conclude that the central star is a M3-M5e late type dwarf. Our medium resolution spectra do not show evidence for the LiI 6708 Å doublet (the equivalent width being less than 0.01 Å (0.04 Å) if the velocity is –135 km/s (0 km/s)). However, lithium is efficiently destructed in low mass stars and consequently the lack of Li absorption only tells us that the star is older than $\approx 10^7$ yr (Rebolo, 1991 and references therein). Radial velocity variations are poorly constrained from our data. However, the medium resolution spectra obtained at Calar Alto (see Fig. 4) seem to exhibit a significantly different H$\alpha$ radial velocity from the bulk of the low resolution spectra which might reflect rotation or orbital motion in a close binary (see Table 1). From our four R band images collected between 1991 November 10 and 12 we put an upper limit of 0.3 mag on the photometric variability of the bright central object.

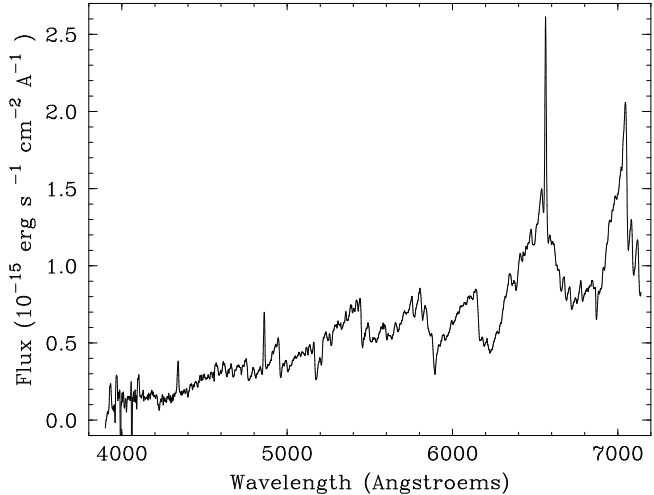

Fig. 3. Mean low resolution spectrum of the central star obtained with the 3.5m telescope at Calar Alto. Spectral features indicate a M3-M5 Ve spectral type.

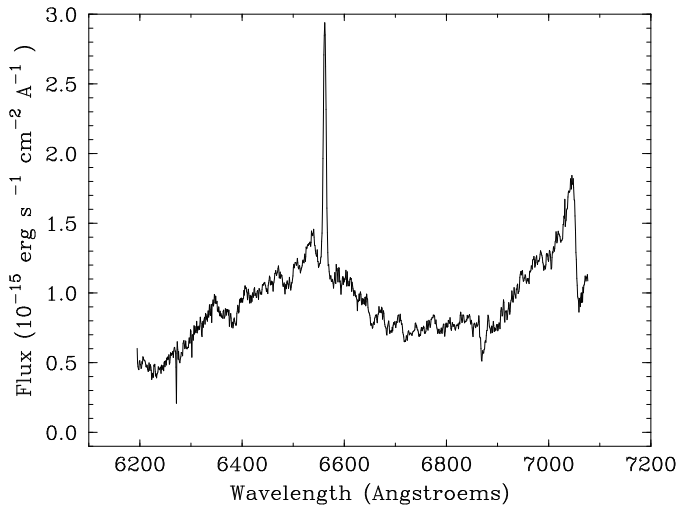

Fig. 4. Mean medium resolution red spectrum of the central star obtained with the 3.5m telescope at Calar Alto. No Li line is detected, indicating an age larger than $10^7$ yrs.

Comparison with photometric standards in NGC 7790 (Christian et al. 1985) yielded B = 18.50 ± 0.20 for the bright central dMe star.

Table 1. Equivalent widths and radial velocities of the dMe star S 10933

| Date of 1992 | UT mid exposure | line used | Rv km/s | error km/s | EW (Å) |
|---|---|---|---|---|---|
| 04:02 | 19:07 | Hα | +52 | 73 | 7 |
| 07:02 | 18:32 | Hβ | +20 | 21 | - |
| 08:02 | 18:47 | Hα | +13 | 63 | 11 |
| 28:09 | 23:39 | Hα | +69 | 46 | 10 |
| 29:09 | 01:48 | Hα | +1 | 46 | 11 |
|  |  | Hβ | −37 | 74 | 12 |
| 30:09 | 00:09 | Hα | −135 | 27 | 10 |
| 30:09 | 01:19 | Hα | −50 | 27 | 10 |
| 30:09 | 03:57 | Hα | −135 | 27 | 10 |
| 01:10 | 01:02 | Hα | +38 | 45 | 13 |
|  |  | Hβ | +27 | 74 | 14 |

### 3.2. Faint companions

Two additional objects lie within or slightly outside the $3\sigma$ error circle of the archival flare event (see Fig. 1). These are fainter than the dMe star by at least two magnitudes. They are aligned southeastwards from the central dMe star together with a third one which is already outside the $3\sigma$ position. Our long slit spectra of the three aligned objects obtained with the 3.5m telescope at Calar Alto show that the brightest notch (object D) located at the end of the apparent alignment is also an M star, however without noticeable Balmer emission (see Fig. 5). No usable spectra could be obtained from notches B and C located in between the dMe star and object D. These objects exhibit flat continua without recognizable emission or absorption features.

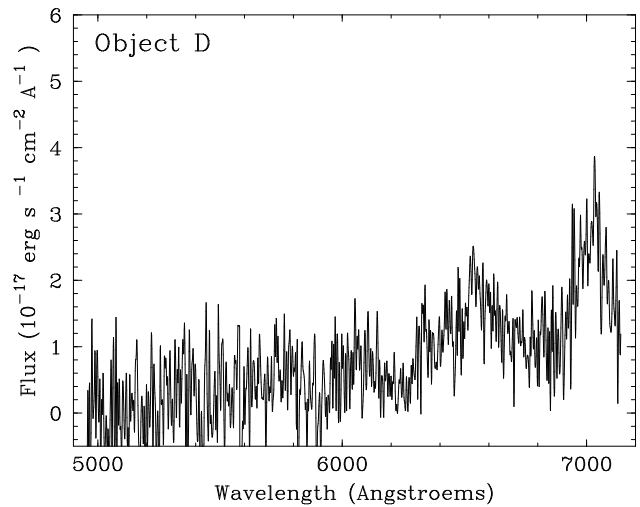

Fig. 5. Mean low resolution spectrum of the brightest notch located at the end of the apparent alignment (object D) starting from the dMe star. TiO molecular bands indicate a M spectral type without noticeable Balmer emission.

**Table 2.** Archival flux limits for S 10933 at other wavelengths

| Date of observation | Detector | Wavelength region | 3σ upper limit |
|---|---|---|---|
| Sep. 1972–Oct. 1976 | Radio | 1420 MHz | 40 Jy* |
| June 27, 1983 | IRAS | 12 μm | 67 mJy |
| and | IRAS | 25 μm | 73 mJy |
| July 8, 1983 | IRAS | 60 μm | 64 mJy |
|  | IRAS | 100 μm | 132 mJy |
| Dec. 21, 1990 | ROSAT | 0.1–2.4 keV | 0.012 cts/sec |

* Reich & Reich (1986 and 1988).

## 4. Discussion

### 4.1. The flare star

Our photometric and spectroscopic observations leave no doubt that the bright dMe star is responsible for the large optical flare recorded in 1966. The spectral type indicates an absolute magnitude $M_V$ in the range of 10 to 14 and a distance of 40 to 250 pc. At a galactic latitude of $b = -48°.6$ ($l = 108°.6$) this corresponds to a height below the galactic plane of the same order. The apparent magnitudes during the flare and in quiescence translate into visual luminosities (blue band) of $4 \times 10^{31}$ (d/100 pc)$^2$ erg/s and $3.5 \times 10^{29}$ (d/100 pc)$^2$ erg/s, respectively.

All available data as listed in Table 2 are consistent with a genuine dMe star. The M flare star coronal X-ray luminosity is well correlated with the bolometric luminosity (Pallavicini et al. 1990) and is a good diagnostics of flare activity. Using this correlation we expect the quiescent X-ray luminosity to be in the range of $7 \times 10^{27}$ to $1 \times 10^{29}$ erg/s. Also the Hα luminosity of $1.6 \times 10^{28}$ (d/100 pc)$^2$ erg/s suggests a quiescent X-ray luminosity of $\approx 1 \times 10^{29}$ (d/100 pc)$^2$ erg/s (Fleming et al. 1988). Using a thermal line + continuum spectrum and the maximum absorbing column in the flare star direction of $N_H = 5 \times 10^{20}$ cm$^{-2}$ the upper limit of 0.012 cts/sec derived from the ROSAT All-Sky-Survey corresponds to $< 2.5 \times 10^{-13}$ erg/cm$^2$/s or $< 3 \times 10^{29}$ (d/100 pc)$^2$ erg/s (in the 0.1–2.4 keV range), consistent with the above estimates.

The blue flare colour derived from the simultaneously exposed blue and visual archival discovery plates is consistent with fast photometric multicolour observations of flare stars (see e.g. Gurzadyan 1980 and references therein) and is in accordance with flare theories (see e.g. Haisch et al. 1991).

The observed temporal profiles of flares vary considerably. However, they are traditionally divided into two classes: the "spike flares" reach their peak intensity within less than a minute while "slow flares" take several minutes for the rise time. The decay to below the half-maximum occurs within minutes, but afterwards most flares take several hours to relax fully to quiescence. Assuming that the flare lasted as long as (or longer than) the exposure time of the discovery plates (40 min), implies a peak amplitude of about $\triangle m_B \approx 5.2$ mag. Alternatively, if the event belongs to the class of fast flares with typical duration shorter than $\approx 10$ minutes, the peak $m_V$ amplitude could have been as high as 7–9 mag. Other large amplitude flaring stars are AF Psc with $\triangle m_B \approx 7.5$ mag (Bond 1976, Greenstein 1977), UV Cet with $\triangle m_B \approx 6.5$ mag (Jarrett & Gibson 1975), EV Lac with $\triangle m_B \approx 6$ mag (Roizman & Shevchenko 1982), the M dwarf secondary of the AM Her binary with $\triangle m_B \approx 5.9$ mag (Shakhovskoy et al. 1993), AD Leo with $\triangle m_B \approx 4.1$ mag (Hawley & Pettersen 1991), and the presumed flare star CZ Cnc with a possible amplitude of $\triangle m_B \approx 9.5$ mag (Lovas 1977, Schaefer 1990). Even a peak amplitude of 5.2 mag and a lower limit for the corresponding flare energy release of $10^{35}$ (d/100 pc)$^2$ erg in the optical bandpass are rare for a flare star and we can conclude that the event witnessed in 1966 was certainly unusual.

### 4.2. The relation to GBS 0008+13

The presumption at the time of the discovery (Greiner et al. 1991) was that the recorded optical flash was a likely counterpart candidate for a GRB. Indeed, this optical flash is well positioned within a 3200 arcmin$^2$ error box of a GRB which occured on October 6, 1978 (GB 781006B = GBS 0008+13, Atteia et al. 1987). The rather extreme value for the flare amplitude might favour speculations whether or not this flare is indeed related to a GRB. In fact, the possible relation with stellar flares was among the first models proposed to explain GRBs (Stecker & Frost (1973), Brecher & Morrison (1974), Karitskaya (1975)). Nearby flare stars are isotropic in the sky and have a -3/2 slope in their logN–logS distribution (for a sampling distance smaller than $\approx$300 pc) and thus meet one of the statistical GRB criteria.

Vahia & Rao (1988) have made detailed correlations with all types of magnetic active stars (flare stars, RS CVn, cataclysmic variables) and concluded that GRBs might be considered as stellar equivalents of solar hard X-ray bursts. Assuming that an increasing amplitude is accompanied by the emission of increasingly higher photon energies, one could expect flare stars of very large amplitude to be responsible for a fraction of the observed

GRBs. It has been argued (Harutyunian & Hayrapetyan 1990) that magnetically active dwarfs with deep convection zones and low absolute magnitudes can be effective $\gamma$-ray sources (in the framework of the so-called pinch model of stellar flares). Recently, Liang and Li (1993) have shown that flare stars could account for as many as half of the BATSE (the Burst and Transient Source Experiment onboard the Compton Gamma-ray Observatory) triggers without violating the observed log N–log S distribution of GRBs. This was based on a sampling distance of $\approx 100$ pc and uses an assumed $L_X/L_{opt} \approx 100$ to extrapolate from the observed optical flare star fluences to the expected hard X-ray fluences at BATSE energies.

Estimating the a priori probability of discovering such a large amplitude flaring star in a rather large (3200 arcmin$^2$) $\gamma$-ray error box is difficult because of the various statistical biases in flare star catalogs and flare observational data bases. There are convincing evidences that flare activity preferentially occurs in rather young stars (e.g. Mirzoyan 1990). Most flaring M stars exhibit strong Balmer emission (Pettersen 1991) while kinematical studies of dMe stars show that they belong to the young disk population with a mean age of $\approx 2 \times 10^9$ yrs (Giampapa & Liebert 1986). However, old stars in close binaries may also exhibit strong flaring activity (Haisch et al. 1991).

Using the stellar population model of Robin & Crézé (1986) which explicitly handle the age parameter we predict that $\approx 10$ stars with B $\leq$ 18.5 and younger than $10^9$ yrs should be present inside the GRB error box, among which two late M type dwarf stars. Therefore, in the absence of stronger constraints on the age of the Me dwarf and reliable statistics on flare amplitudes, we cannot prove that this particular flare star has caused the $\gamma$-ray burst GB 781006B neither can we dismiss the flare star explanation for this particular GRB.

### 4.3. Stellar flares as background events for GRB optical counterpart searches

The known flare stars (with typical flare amplitudes of 0.5–2 mag) have primarily been discovered by patrols reaching 12–15th mag. Optical monitoring is much more sparse at fainter magnitudes, however. Thus, the selection effect against low luminosity, large amplitude flares is twofold: 1) large amplitude flares are hard to detect because they are less frequent than small amplitude flares. 2) since large amplitude flares predominantly originate from faint stars, many of these may have remained unknown either due to their faint luminosity with respect to optical patrols or due to the previous practice of variable star researchers to often ignore unique optical archival objects which have no faint (e.g. POSS) counterpart (thus interpreting these as plate defects). Thus, it is conceivable that low luminosity flare stars are more common than previously thought as was noted already earlier (Schaefer 1990).

Since there exist no reliable distributions of flare amplitudes (at least for amplitudes larger than 2 mag) a crude estimate of this number can be obtained in two ways: First, collecting data from many solar neighbourhood flare stars and those in the Orion and Pleiades clusters, Gershberg (1989) reviewed the total energy output in the blue band $E_B$ per flare in dependence of flare frequency ($\nu$). According to this, the frequency of flares with $10^{34}$ erg energy release is of the order of $10^{-2.5...-3}$ h$^{-1}$ for solar neighbourhood and $10^{-3.5...-4.5}$ h$^{-1}$ for cluster flare stars. The difference between the flash frequencies in local field and cluster flare stars might be either due to the different age of the stars or due to the completely different detection methods which favor the detection of long flares (large energy output) for cluster stars. Thus, we can only state that either our object has one of the largest energy output of any observed local flare star with a frequency of the order of $10^{-2.5...-3}$ h$^{-1}$, or our object has a property as the cluster flare stars so that the energy output is typical but the frequency is only $10^{-3.5...-4.5}$ h$^{-1}$. If the lower flare frequency would be due to different ages, the latter might be a hint toward a young dMe star. Second, using the flare amplitude distribution of the $\eta$ Tau field (Szecsenyi-Nagy 1990) and transforming the relative flare numbers into flare frequencies with respect to the 0.8–1.0 mag amplitude bins, results in $10^{-4.3}$ h$^{-1}$. This is consistent with the above estimates of cluster flare star frequencies.

The unusually large optical amplitude of the flare is likely to be linked to the faint absolute magnitude of the late type star ($M_V = 10$–14), the rather large monitored sky area (3200 $'^2$) and to the long monitoring time (1375 h) available from the plate collection. This obviously favours the discovery of rare events. The fact that no other flare was found during this monitoring time at Sonneberg with a typical limiting magnitude of $m_{pg} \approx 14$ restricts the frequency of other flares from this star to less than $7 \times 10^{-4}$ h$^{-1}$ for amplitudes greater than 4.5 mag.

Though one seems to know the maximum energy per flare for the Sun and very few flare stars (from their flattening in the log $E_B$ – log $\nu$ diagram), there is no hint for a general maximum flare energy so far. This means, that flares with larger energy output (and thus, in general, larger amplitude) should occur and will certainly be detected with increasing monitoring time.

This makes it difficult to prove or disprove a flare star origin for any optical flash found in the search for $\gamma$-ray burst counterparts. There is basically no hope of identifying a large-amplitude flare star via observed repetitions of similar flares due to the huge amount of monitoring time which would be necessary. Furthermore, if an optical flare is observed from an object fainter than, say $\approx 24$ mag in quiescence, then any spectroscopic proof of a flare star origin is presently rather difficult to achieve. Finally, also the positional coincidence with a GRB error box cannot be used as a (statistical) argument unless the error box is very small (less than $\approx$1 arcmin$^2$). Thus, for the identification of a possible GRB counterpart at optical wavelengths only simultaneous observations could provide some easily conclusive confidence.

## 5. Summary

We have identified the 1966 optical flash event to be caused by a B = 18$^{\rm m}$5 mag dMe flare star. We believe that the relation between this flare star and the $\gamma$-ray burst source GBS 0008+13 is possible but rather unlikely. Even in this case of an unique archival object observed on synchronously exposed plates inside a GRB error box we are still left with an explanation other than a $\gamma$-ray burst optical counterpart.

Our identification confirms the feasibility of searching archival plate collections for real optical flashes of astrophysical origin. Unfortunately, the existence of an unknown number of such large amplitude flaring stars as the one reported here poses serious problems for wide-field GRB optical flash searches.

*Acknowledgements.* We thank H. Steinle for considerable support during the spectroscopic observations at Calar Alto, and W. Wenzel and R. Gershberg for extensive discussions on the observational evidence of large amplitude flaring stars. We greatly acknowledge help with the IRAS data by J. Melisse, Leiden, and thank A. Robin for providing detailed stellar population modelling. Also, we thank the referee, R. Pallavicini, for suggesting several clarifications and improvements. JG was partly supported by the Deutsche Agentur für Raumfahrtangelegenheiten (DARA) GmbH under contract number FKZ 50 OR 9201. CM acknowledges support from a CNRS-MPG cooperation contract.